\documentclass[journal]{IEEEtran}
\usepackage{amsmath,amssymb,bm}
\usepackage{graphicx}
\usepackage{booktabs}
\usepackage{array}
\usepackage{cite}
\usepackage{xcolor}
\usepackage[colorlinks=true,citecolor=blue,linkcolor=blue,urlcolor=blue]{hyperref}

\newcommand{\CN}{\mathcal{CN}}

\title{Noise-Domain OTFS: Statistical Delay--Doppler Signaling for Future Wireless Networks}

\author{Mahmoud~Aldababsa,~\IEEEmembership{Senior Member,~IEEE}%
\thanks{Mahmoud Aldababsa is with the Department of Electrical and Electronics Engineering, Faculty of Engineering and Architecture, Istanbul Nisantasi University, Istanbul, T\"urkiye.}}

\begin{document}
\maketitle

\begin{abstract}
Orthogonal time frequency space (OTFS) usually conveys information through deterministic constellation symbols placed on a delay--Doppler (DD) grid. This article develops a different viewpoint, termed \emph{noise-domain OTFS} (ND-OTFS), in which each message selects the covariance of a random DD-domain vector. The resulting statistical codeword can encode information through per-bin variance, equal-energy correlation structure, or separable delay--Doppler covariance. We explain the transceiver principle, receiver implications, reliability--rate tradeoffs, and practical challenges. Representative results show that a matched-payload coherent OTFS reference retains a clear raw-reliability advantage, whereas ND-OTFS enables information recovery from covariance geometry even when competing states have identical average energy. Additional statistical observations reduce finite-sample overlap at the cost of latency and rate. ND-OTFS therefore complements, rather than replaces, deterministic OTFS by introducing a matrix-valued statistical signaling dimension across DD resources.
\end{abstract}


\section{Why Rethink Signaling in the Delay--Doppler Domain?}
Orthogonal time frequency space (OTFS) modulation has attracted considerable interest for high-mobility and doubly selective channels because it represents the propagation environment in the delay--Doppler (DD) domain and spreads information across the available time--frequency resources \cite{Hadani2017OTFS,Raviteja2018OTFS,10584089}. In most OTFS systems, however, the DD grid is still populated by familiar deterministic constellation symbols, such as QAM or PSK. The waveform representation changes, but the underlying information-bearing object remains an instantaneous complex symbol.

Noise-domain communication suggests a different possibility. Thermal-noise communication and the broader NoiseMod framework have shown that information can be carried by the statistical properties of random waveforms rather than by deterministic amplitudes and phases \cite{Basar2023TherCom,Basar2024NoiseMod}. Subsequent designs have explored on--off noise signaling, mean/variance alphabets, multiuser noise-domain signaling, spread-spectrum noise modulation, and differential noise-domain structures \cite{Anjos2025OODN,Anjos2026OODN,Yapici2025NDNOMA,Zayyani2026SSNM,Zayyani2026Composite,Tome2026DBN}. These developments motivate a natural question for OTFS: instead of placing deterministic symbols on the DD grid, can the DD grid itself be used as a structured statistical signaling space?

\begin{figure*}[t]
\centering
\includegraphics[width=0.98\textwidth]{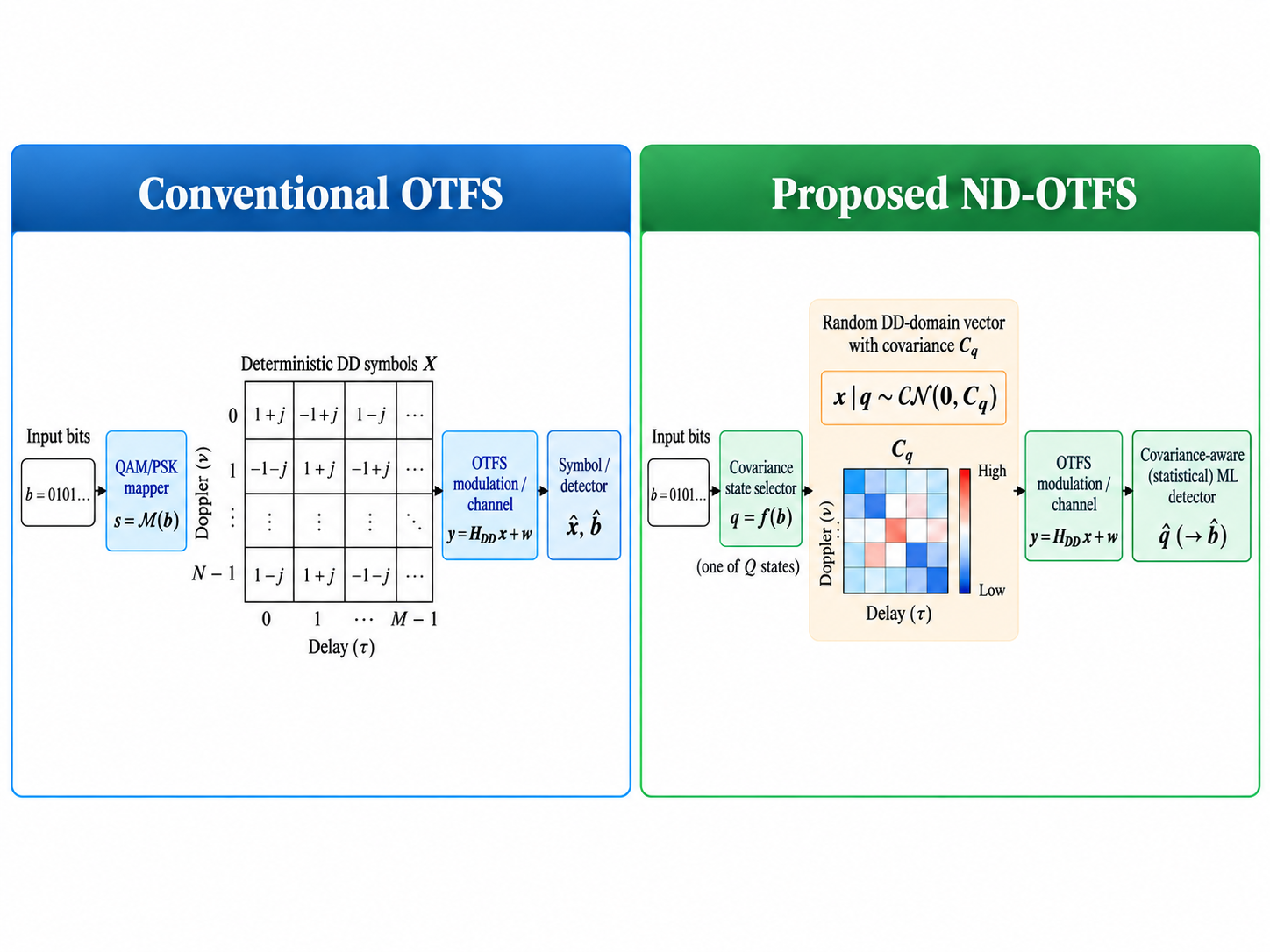}
\caption{Conceptual comparison of conventional OTFS and ND-OTFS signaling. Conventional OTFS maps input bits to deterministic delay--Doppler symbols, whereas ND-OTFS maps the message to a covariance state and generates a random delay--Doppler vector whose second-order statistical structure carries the information.}
\label{fig:concept}
\end{figure*}

ND-OTFS answers this question by making the covariance of the DD-domain transmit vector an information-bearing quantity. The distinction is illustrated in Fig.~\ref{fig:concept}. Conventional OTFS maps input bits to deterministic DD symbols and performs symbol recovery at the receiver. ND-OTFS maps the message to a covariance state, generates a random DD vector according to that state, and performs covariance-aware statistical detection. The instantaneous DD samples therefore need not repeat from one transmission to the next even when the same message is sent.

This shift in viewpoint is more than a replacement of QAM by random samples. If every DD bin were independently assigned only a low or high variance, the result would be a direct extension of scalar NoiseMod. The distinctive opportunity in OTFS is the two-dimensional DD geometry: information can be embedded not only in per-bin variances, but also in deliberately designed correlations across delay bins, Doppler bins, or both. Thus, the DD grid becomes a structured statistical codeword rather than merely a container for instantaneous symbols. The central novelty is therefore not the use of random DD samples itself, but the treatment of the \emph{matrix-valued covariance structure across delay and Doppler resources as an information-bearing object}.

\section{The Core ND-OTFS Principle}
Let an OTFS frame contain $M$ delay bins and $N$ Doppler bins, so that $D=MN$. After vectorizing the DD grid, denote the transmit vector by $\mathbf{x}\in\mathbb{C}^{D}$. In ND-OTFS, a message index $q$ selects a positive-semidefinite covariance matrix $\mathbf{C}_q$, and the DD vector is generated as
\begin{equation}
\mathbf{x}\mid q\sim\CN(\mathbf{0},\mathbf{C}_q).
\label{eq:tx}
\end{equation}
The message therefore determines the second-order statistics of the transmitted DD vector, while the instantaneous realization remains random.

The random DD-domain vector then propagates through the effective OTFS DD-domain channel and is corrupted by receiver noise. Conditioned on $q$ and the channel, the received vector remains Gaussian with covariance
\begin{equation}
\boldsymbol{\Sigma}_q
=\mathbf{H}_{\rm DD}\mathbf{C}_q\mathbf{H}_{\rm DD}^{H}+N_0\mathbf{I}.
\label{eq:rx-cov}
\end{equation}
Hence, the channel does not destroy the statistical codeword; rather, it maps the selected transmit covariance into a message-dependent received covariance.

A channel-aware covariance detector evaluates how well the received DD vector matches each candidate covariance model and selects the most likely state. For equiprobable states, this is the standard Gaussian maximum-likelihood decision based jointly on the covariance volume and a covariance-weighted quadratic compatibility measure. Importantly, the receiver does not need to reconstruct the instantaneous random transmit realization before deciding the information state. This statistical receiver interpretation is illustrated in Fig.~\ref{fig:three-types}; the exact analytical detector and its error analysis are left to the companion technical treatment so that the present article remains tutorial in emphasis.

\section{Three Ways to Encode Information in DD Statistics}
\begin{figure*}[t]
\centering
\includegraphics[width=0.98\textwidth]{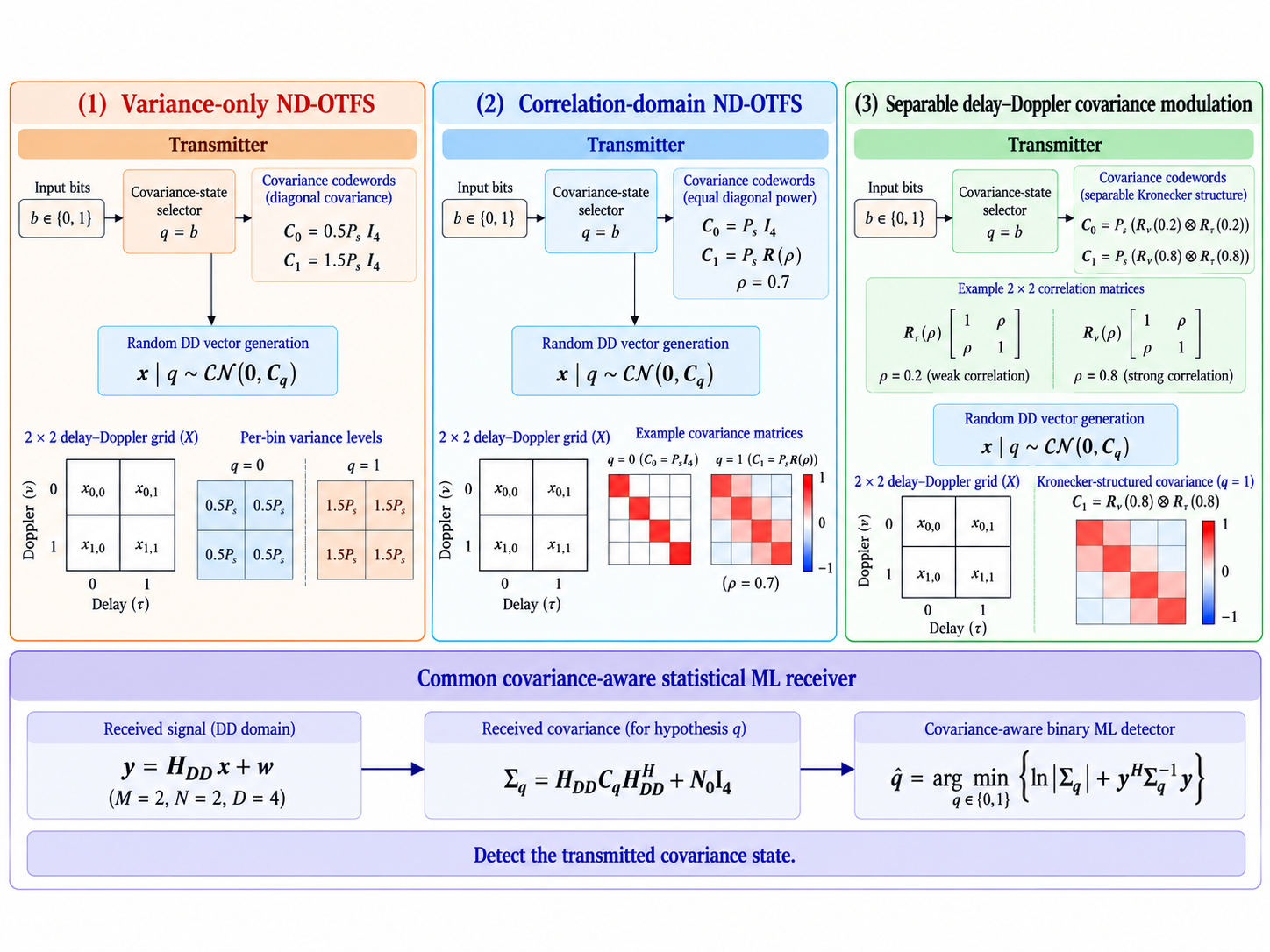}
\caption{Three representative binary ND-OTFS realizations: variance-only, correlation-domain with fixed marginal power, and separable delay--Doppler covariance signaling.}
\label{fig:three-types}
\end{figure*}

The same ND-OTFS architecture supports several covariance alphabets. Fig.~\ref{fig:three-types} summarizes three representative binary realizations through a pedagogical $M=2$, $N=2$ example. The small grid is used only to make the covariance structure visually transparent; practical frames can be much larger.

\subsection{Variance-Only ND-OTFS}
The simplest construction uses diagonal covariance matrices. Different information states are distinguished through different variance levels assigned to the DD bins. This mode is conceptually closest to conventional variance-based noise modulation and can be attractive when transmitter simplicity is the main objective. It also provides a useful baseline because the receiver can exploit multiple DD samples associated with the same statistical state.

The limitation is equally clear: if only diagonal power levels are used, the two-dimensional geometry of the DD grid is not fully exploited. The information remains tied mainly to energy statistics.

\subsection{Correlation-Domain ND-OTFS}
A more distinctive mode fixes the marginal power of every DD component while changing the off-diagonal correlations. For example, one binary state may use an identity covariance, while the competing state uses a correlation matrix with the same diagonal and the same trace. Both states then have identical average transmit energy, yet they induce different DD-domain dependence patterns.

This case captures the central conceptual advantage of covariance signaling: information can reside in \emph{how} random DD samples are related, rather than in how much average energy they contain. Such a construction may be useful when large state-dependent power changes are undesirable or when statistical structure is itself an attractive design resource.

\subsection{Separable Delay--Doppler Covariance Signaling}
The third realization explicitly exploits the two axes of the DD grid. A structured covariance codeword can be formed from a delay-correlation matrix and a Doppler-correlation matrix through a Kronecker product. This yields a compact parameterization with a direct physical interpretation: one set of parameters controls imposed dependence across delay bins, while another controls dependence across Doppler bins.

This separable structure is attractive for two reasons. First, it reduces the number of covariance parameters that must be stored and optimized. Second, it preserves the uniquely two-dimensional character of OTFS, making the statistical alphabet easier to interpret than an arbitrary dense covariance codebook.

\section{What Changes at the Receiver?}
The three ND-OTFS realizations differ mainly in the statistical feature used to convey information. As summarized in Table~\ref{tab:modes}, variance-only signaling relies on diagonal power variations, correlation-domain signaling embeds information in off-diagonal dependence while preserving the marginal power, and separable DD signaling jointly controls the delay- and Doppler-domain correlation structure. These alternatives provide different tradeoffs among implementation simplicity, statistical separability, and exploitation of the two-dimensional DD geometry.

\begin{table}[t]
\caption{Qualitative Comparison of the Three ND-OTFS Realizations}
\label{tab:modes}
\centering
\begin{tabular}{p{0.23\columnwidth}p{0.26\columnwidth}p{0.39\columnwidth}}
\toprule
Mode & Information-bearing feature & Main implication \\
\midrule
Variance-only & Diagonal variances & Simplest implementation; mainly energy-domain distinction \\
Correlation-domain & Off-diagonal dependence with fixed diagonal & Can separate equal-average-energy states through correlation geometry \\
Separable DD & Delay and Doppler correlation factors & OTFS-specific structure with reduced design/storage dimension \\
\bottomrule
\end{tabular}
\end{table}

Regardless of the selected covariance alphabet, the receiver operates on the same basic principle: it determines which candidate statistical state most plausibly explains the observed DD-domain vector. Consequently, the relevant notion of separation is no longer the Euclidean distance between deterministic constellation points, but the distinguishability between the corresponding received distributions.

This difference has several consequences. First, statistical distances between received distributions become more informative than Euclidean distance between constellation points. Second, the receiver can operate without reconstructing the random transmit realization itself. Third, receiver performance depends on the observation dimension because covariance discrimination is inherently statistical: a larger DD vector, or multiple independent observations of the same state, provides more evidence about the underlying covariance.

At the same time, ND-OTFS is not automatically noncoherent. The covariance-aware receiver assumes knowledge of the effective DD channel through the received-covariance model in \eqref{eq:rx-cov}. Channel acquisition therefore remains an important practical issue. Noise-domain channel estimation has already been studied in simpler random-waveform settings, where pilot-assisted estimators can operate even though the instantaneous transmitted noise samples are unknown \cite{Shen2025Channel}. Extending such ideas to a full OTFS DD-channel matrix is an open design problem rather than a solved component of the present framework.

\section{Reliability, Rate, and Observation Length}
Statistical signaling introduces a tradeoff that does not appear in the same form for deterministic constellations. Two full-rank Gaussian covariance states have overlapping probability densities. Consequently, one finite-dimensional observation can be misclassified even when additive receiver noise becomes very small. In other words, the limiting reliability is governed not only by thermal noise but also by intrinsic distribution overlap.

There are two direct ways to improve discrimination. The first is to increase the separation between covariance states. For correlation-domain signaling, for example, stronger differences between the identity state and the correlated state generally make the induced received distributions easier to distinguish. The second is to increase the amount of statistical evidence, either by using a higher-dimensional DD observation or by observing the same state over several independent snapshots.

These improvements are not free. Repeating the same covariance state over multiple snapshots consumes observation time and reduces effective information rate. Likewise, using one covariance state over an entire large DD frame may yield excellent statistical discrimination but convey only a small number of information bits per frame. A practical system can address this by partitioning the DD grid into groups and assigning an independent covariance state to each group, thereby trading statistical observation dimension against payload.

This reliability--rate tension is one of the most important system-level questions for ND-OTFS. In particular, a binary covariance state applied to an entire $M\times N$ frame carries only one bit, so whole-frame signaling is best interpreted as a proof of concept rather than a spectrally efficient operating point. A practical extension is to partition the DD grid into groups and assign an independent covariance state to each group, or to use higher-order covariance alphabets. Future designs should therefore jointly choose group size, alphabet order, observation length, and covariance separation instead of optimizing error probability alone.

\section{Representative Numerical Insights}
A fair reliability comparison must separate the signaling principle from the large payload difference between whole-frame binary covariance signaling and conventional full-rate OTFS. We therefore use a \emph{matched-payload, matched-energy coherent OTFS benchmark}: one bit selects one of two deterministic DD codewords, while each ND-OTFS mode conveys the same one bit/frame with the same average frame energy, DD observation dimension, channel realization, and channel knowledge. This comparison isolates raw binary reliability; it is \emph{not} a matched-net-spectral-efficiency comparison with full-rate BPSK/QPSK OTFS.

\begin{figure}[t]
\centering
\includegraphics[width=\columnwidth]{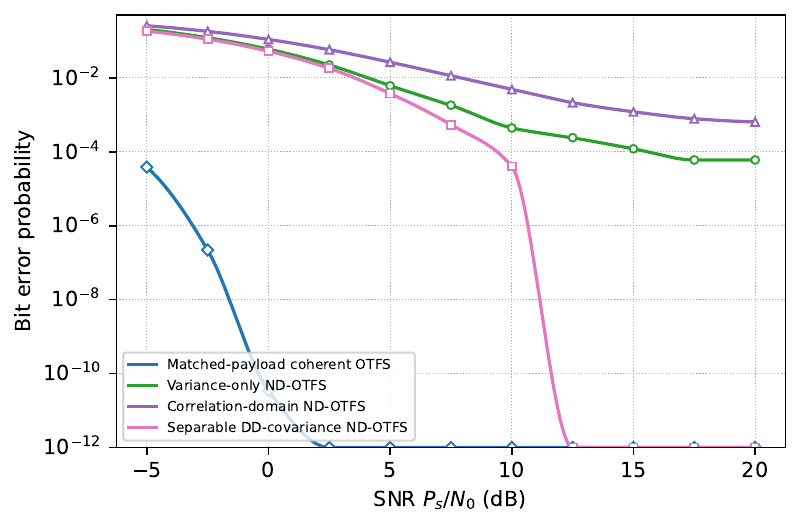}
\caption{Matched-payload reliability comparison between coherent OTFS and the three ND-OTFS modes for an $8\times8$ DD grid. All schemes convey one bit/frame with the same average frame energy and channel knowledge; the comparison isolates raw binary reliability rather than matched net spectral efficiency.}
\label{fig:bep-mag}
\end{figure}

Fig.~\ref{fig:bep-mag} compares the coherent benchmark with the variance-only, correlation-domain, and separable DD covariance modes. The coherent reference has the expected reliability advantage because its deterministic hypotheses become increasingly separable as receiver noise vanishes. By contrast, finite-dimensional Gaussian covariance states retain statistical overlap. The important ND-OTFS observation is different: correlation-domain signaling remains distinguishable even when competing states have identical average energy and identical marginal DD power, confirming that off-diagonal DD dependence can itself carry information. The separable construction further illustrates how covariance-codebook geometry changes statistical distinguishability.

\begin{figure}[t]
\centering
\includegraphics[width=\columnwidth]{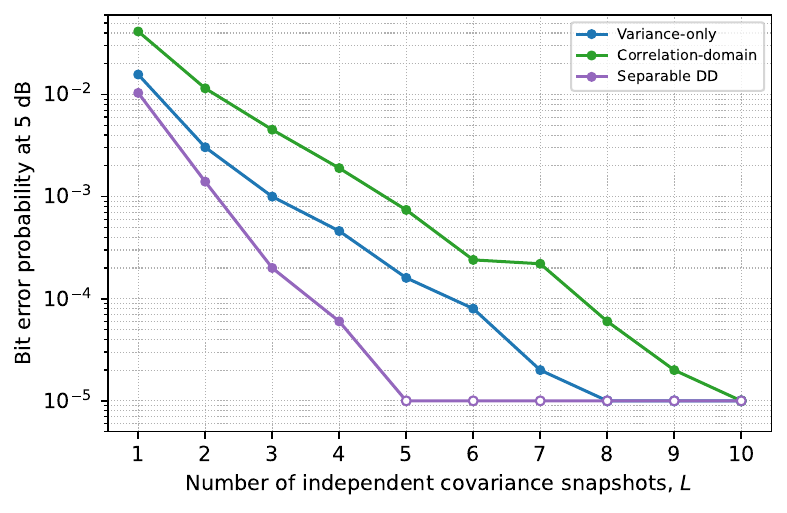}
\caption{Reliability--observation tradeoff at $P_s/N_0=5$ dB. Increasing the number of independent covariance snapshots improves statistical discrimination at the cost of observation time and effective rate.}
\label{fig:snap-mag}
\end{figure}

Fig.~\ref{fig:snap-mag} highlights the complementary reliability--observation tradeoff. Repeating the same covariance state over additional independent observations provides more statistical evidence and lowers the error probability, but it also increases observation time and, without additional multiplexing, reduces effective rate. The snapshot count $L$ is therefore a system-level parameter that jointly affects reliability, latency, and payload density rather than merely a receiver setting.

Together, Figs.~\ref{fig:bep-mag} and \ref{fig:snap-mag} position ND-OTFS as a complementary statistical signaling architecture rather than a universal BER-improving replacement for coherent OTFS. A full system comparison should next evaluate grouped or higher-order ND-OTFS at matched net spectral efficiency, coding, pilot/guard overhead, latency, bandwidth, and implementation energy.

\section{Where Could ND-OTFS Be Useful?}
ND-OTFS should not be viewed as a universal replacement for QAM/PSK OTFS. Its value is more likely to emerge in scenarios where statistical waveform generation or covariance-domain processing provides a system advantage.

\subsection{Low-Power and Low-Complexity End Devices}
Noise-domain communication has been motivated in part by the possibility of generating information-bearing random waveforms with very simple circuitry \cite{Basar2023TherCom}. If practical RF implementations can generate controlled DD covariance states efficiently, ND-OTFS could become an alternative signaling mode for devices whose transmitter complexity or energy budget is more restrictive than their receiver-side processing capability. This is a research opportunity rather than a demonstrated energy advantage of the present proof of concept.

\subsection{High-Mobility Links}
OTFS is naturally associated with channels exhibiting substantial Doppler variation. ND-OTFS preserves the DD-domain channel representation while changing the information-bearing object from deterministic symbols to covariance states. This makes high-mobility communication a natural environment in which to investigate statistical DD structures. The present proof of concept does not establish a high-mobility robustness advantage; fractional Doppler, practical pulses, and synchronization effects must be incorporated before such a claim can be assessed.

\subsection{Sensing-Aware and Statistical Waveforms}
A covariance-shaped random waveform can simultaneously possess communication and sensing-relevant statistical structure. Although a joint communication-sensing design is beyond the present article, covariance selection creates a natural interface for future work in which statistical DD patterns are optimized for both message separation and sensing objectives.

\subsection{Hardware-Constrained or Spectrally Shaped Transmission}
Structured covariance alphabets can impose spectral or correlation properties by design. This may be useful when the transmitter must satisfy hardware, coexistence, or spectral-shaping constraints while still embedding information in second-order statistics.

\section{Open Challenges and Research Roadmap}
Several issues must be resolved before ND-OTFS can progress from a mathematical framework to a practical signaling architecture.

\emph{Channel estimation and mismatch:} the covariance-aware receiver depends on the effective DD channel. Practical designs must estimate this channel from random training signals and quantify how estimation error affects covariance-state decisions.

\emph{Fractional Doppler and practical pulses:} the present proof of concept uses an integer-grid DD model to isolate the signaling principle. Fractional Doppler, practical pulse shapes, inter-Doppler interference, synchronization errors, and RF front-end effects should be incorporated explicitly.

\emph{Higher-order and grouped statistical alphabets:} whole-frame binary covariance states have very low payload density. Structured $Q$-ary alphabets, DD grouping, labeling, and coding are needed to approach practical spectral efficiencies while preserving enough samples for reliable covariance discrimination.

\emph{Rate--complexity--reliability co-design:} larger observation groups and more snapshots improve statistical separation but increase receiver work and reduce payload density or increase latency. Group size, observation length, training overhead, and detector complexity should therefore be optimized jointly.

\emph{Covariance synthesis hardware and fair system baselines:} mathematical covariance states must ultimately be produced by realizable RF hardware, with calibration, quantization, power-amplifier, and energy-consumption effects included. Comparisons with conventional OTFS should then be performed at matched net spectral efficiency, coding, bandwidth, latency, and implementation energy.

A practical roadmap has three stages: analytical characterization under realistic OTFS channels, system-level co-design of grouping, rate, training, complexity, and energy, and experimental validation of covariance-state generation and detection on software-defined-radio or RF testbeds. The broader objective is a family of DD-domain waveforms whose two-dimensional covariance geometry is deliberately designed to carry information.

\section{Conclusion}
ND-OTFS extends noise-domain communication from scalar statistical parameters to structured random vectors over the OTFS DD grid. Instead of mapping a message only to deterministic QAM/PSK symbols, the transmitter selects a covariance state whose variance and/or delay--Doppler correlation geometry carries information. A covariance-aware receiver then discriminates the corresponding received distributions. The framework exposes a new reliability--rate--complexity tradeoff: deterministic coherent OTFS retains a clear raw-reliability advantage in the matched-payload example, whereas ND-OTFS offers an additional statistical information-bearing dimension, including equal-energy correlation signaling. The next steps are higher-rate grouped alphabets, realistic DD channels and channel estimation, matched-rate system comparisons, and hardware validation.

\bibliographystyle{IEEEtran}
\bibliography{references}

@INPROCEEDINGS{Hadani2017OTFS,
  author={Hadani, R. and Rakib, S. and Tsatsanis, M. and Monk, A. and Goldsmith, A. J. and Molisch, A. F. and Calderbank, R.},
  booktitle={2017 IEEE Wireless Communications and Networking Conference (WCNC)}, 
  title={Orthogonal Time Frequency Space Modulation}, 
  year={2017},
  volume={},
  number={},
  pages={1-6},
  doi={10.1109/WCNC.2017.7925924}}

@ARTICLE{Raviteja2018OTFS,
  author={Raviteja, P. and Phan, Khoa T. and Hong, Yi and Viterbo, Emanuele},
  journal={IEEE Transactions on Wireless Communications}, 
  title={Interference Cancellation and Iterative Detection for Orthogonal Time Frequency Space Modulation}, 
  year={2018},
  volume={17},
  number={10},
  pages={6501-6515},
  doi={10.1109/TWC.2018.2860011}}

@ARTICLE{10584089,
  author={Aldababsa, Mahmoud and \"{O}zyurt, Serdar and Kurt, G\"{u}ne\c{s} Karabulut and Kucur, O\u{g}uz},
  journal={IEEE Open Journal of the Communications Society}, 
  title={A Survey on Orthogonal Time Frequency Space Modulation}, 
  year={2024},
  volume={5},
  number={},
  pages={4483-4518},
  doi={10.1109/OJCOMS.2024.3422801}}

@article{Basar2023TherCom,
  author={Basar, Ertugrul},
  journal={IEEE Transactions on Communications}, 
  title={Communication by Means of Thermal Noise: Toward Networks With Extremely Low Power Consumption}, 
  year={2023},
  volume={71},
  number={2},
  pages={688-699},
  doi={10.1109/TCOMM.2022.3228290}}

@article{Basar2024NoiseMod,
  author={Basar, Ertugrul},
  journal={IEEE Wireless Communications Letters}, 
  title={Noise Modulation}, 
  year={2024},
  volume={13},
  number={3},
  pages={844-848},
  doi={10.1109/LWC.2023.3346471}}

@article{Shen2025Channel,
  author={Shen, Hong and Yang, Zhou and Chen, Yang},
  journal={IEEE Wireless Communications Letters}, 
  title={Channel Estimation via Thermal Noises}, 
  year={2025},
  volume={14},
  number={1},
  pages={178-182},
  doi={10.1109/LWC.2024.3491813}}

@article{Anjos2025OODN,
  author={Anjos, Andr\'{e} A. dos and Silva, Hugerles S.},
  journal={IEEE Wireless Communications Letters}, 
  title={On--Off Digital Noise Modulation}, 
  year={2025},
  volume={14},
  number={11},
  pages={3595-3599},
  doi={10.1109/LWC.2025.3599037}}

@article{Anjos2026OODN,
  author={Anjos, Andr\'{e} A. Dos and Braga, Victor H. F. and Silva, Hugerles S. and Vieira, Robson D.},
  journal={IEEE Wireless Communications Letters}, 
  title={On-Off Digital Noise Modulation: Optimal Likelihood Threshold and Exact BEP in AWGN and {$\alpha$-$\mu$} Fading}, 
  year={2026},
  volume={15},
  number={},
  pages={1474-1478},
  doi={10.1109/LWC.2026.3655586}}

@article{Yapici2025NDNOMA,
  author={Yapici, Erkin and Islam Tek, Yusuf and Basar, Ertugrul},
  journal={IEEE Open Journal of the Communications Society}, 
  title={Noise-Domain Non-Orthogonal Multiple Access}, 
  year={2025},
  volume={6},
  number={},
  pages={8410-8421},
  doi={10.1109/OJCOMS.2025.3615839}}

@article{Zayyani2026SSNM,
 author={Zayyani, Hadi and Salman, Mohammad and de Figueiredo, Felipe A. P. and de Souza, Rausley A. A.},
  journal={IEEE Communications Letters}, 
  title={Spread Spectrum Noise Modulation: Analysis and Detection}, 
  year={2026},
  volume={30},
  number={},
  pages={1106-1110},
  doi={10.1109/LCOMM.2026.3661635}}

@article{Zayyani2026Composite,
  author={Zayyani, Hadi and Salman, Mohammad and Garc\'{i}a, Fernando D. A. and De Figueiredo, Felipe A. P. and De Souza, Rausley A. A.},
  journal={IEEE Communications Letters}, 
  title={Composite Generalized Quadratic Noise Modulation via Signal Addition: Towards Higher-Dimensional Noise Modulations}, 
  year={2026},
  volume={},
  number={},
  pages={1-1},
  doi={10.1109/LCOMM.2026.3724981}}

@article{Tome2026DBN,
  author={Tome, Paulo V. B. and dos Anjos, Andre A. and Silva, Hugerles S. and Araujo, Daniel C. and Vieira, Robson D. and Basar, Ertugrul},
  title={Receive Diversity for Differential Binary Noise Modulation},
  journal={arXiv preprint arXiv:2608.23950},
  year={2026}
}

\end{document}